\documentclass{aa}

\usepackage{graphicx}
\usepackage{amsmath}
\usepackage{amssymb}
\usepackage{xspace}
\usepackage{commath}
\usepackage{times}
\usepackage{bm} 
\usepackage{balance}
\usepackage{hyperref}
\usepackage{mathtools}
\usepackage{stmaryrd}
\usepackage{tabulary}
\usepackage{xcolor}
\usepackage{graphicx}
\usepackage{txfonts}
\usepackage{orcidlink}

\usepackage{afterpage}
\usepackage{multicol}
\newsavebox{\shortpagebox}

\makeatletter
\newcommand{\shortpage}[1]{\par
  \setbox\shortpagebox=\vbox{\strut #1\par}\afterpage{\onecolumn
    \begin{multicols}{2}
    \unvbox\AP@partial
    \end{multicols}}\unvbox\shortpagebox
\par}
\makeatother

\hypersetup{dvips, colorlinks=true, linkcolor=blue, citecolor=blue, filecolor=blue, urlcolor=blue}

\usepackage{soul} \usepackage{xcolor}
\definecolor{aquamarine}{rgb}{0.5, 1.0, 0.83}

\usepackage{hyperref}

\newcommand\soutpars[1]{\let\helpcmd\sout\parhelp#1\par\relax\relax}

\defcitealias{Tep2022}{T22}

\usepackage{acronym}

\newcommand{\half}{\tfrac{1}{2}}
\newcommand{\p}{\partial}
\newcommand{\rd}{\mathrm{d}}
\newcommand{\rr}{\mathrm{r}}
\newcommand{\rc}{\mathrm{c}}
\newcommand{\rp}{\mathrm{p}}
\newcommand{\ra}{\mathrm{a}}
\newcommand{\rt}{\mathrm{t}}

\newcommand{\rrh}{\mathrm{rh}}
\newcommand{\rHU}{\mathrm{HU}}

\newcommand{\rrp}{r_{\rp}}
\newcommand{\rra}{r_{\ra}}

\newcommand{\trh}{t_{\rrh}}

\newcommand{\Rc}{R_{\rc}}
\newcommand{\br}{\boldsymbol{r}}
\newcommand{\bbv}{\boldsymbol{v}}

\newcommand{\vrr}{v_{\rr}}
\newcommand{\vt}{v_{\rt}}

\newcommand{\bJ}{\boldsymbol{J}}
\newcommand{\Jr}{J_{\rr}}
\newcommand{\Lz}{L_z}

\newcommand{\btheta}{\boldsymbol{\theta}}

\newcommand{\Ftot}{F_{\mathrm{tot}}}

\newcommand{\dvPar}{\langle\Delta v_{\parallel} \rangle}
\newcommand{\dvParSq}{\langle (\Delta v_{\parallel})^2 \rangle}
\newcommand{\dvPerpSq}{\langle (\Delta v_{\perp})^2 \rangle}
\newcommand{\dE}{\langle \Delta E \rangle}
\newcommand{\dL}{\langle \Delta L \rangle}
\newcommand{\dESq}{\langle (\Delta E)^2 \rangle}
\newcommand{\dLSq}{\langle (\Delta L^2) \rangle}
\newcommand{\dEL}{\langle \Delta E \Delta L \rangle}

\newcommand{\mF}{\mathcal{F}}
\newcommand{\mbF}{\boldsymbol{\mF}}

\newcommand{\bD}{\boldsymbol{D}}
\newcommand{\tmax}{t_{\max}}

\newcommand{\Nrun}{N_{\mathrm{run}}}

\usepackage{soul} \usepackage{xcolor}
\definecolor{aquamarine}{rgb}{0.5, 1.0, 0.83}

\newcommand{\dchchange}{\color{green}}

\begin{document}

 \title{ Discrepancies between Chandrasekhar's theory of relaxation and $N$-body simulations}

 \authorrunning{K. Tep  \& D. C. Heggie}
 
  \author{
          Kerwann~Tep\inst{1,2}\orcidlink{0009-0002-8012-4048} \and
          Douglas~C.~Heggie\inst{3}\orcidlink{0000-0002-1910-4630}
          }
          
 \institute{
            Observatoire Astronomique de Strasbourg,  CNRS UMR 7550, 11 rue de l’Universit\'e, F-67000 Strasbourg, France  
\and
            Department of Physics and Astronomy, The University of North Carolina at Chapel Hill, Chapel Hill, NC 27599, USA
         \and
             School of Mathematics and Maxwell Institute for Mathematical Sciences, University of Edinburgh, Kings Buildings, Edinburgh EH9 3FD,UK
             }

\abstract{Globular clusters are systems which are known to be particularly well described by two-body relaxation.  In recent decades
many studies have shown that Chandrasekhar's orbit-averaged theory is able to reproduce many features of numerical simulations. However, it has been claimed that differences between the theory and simulation remain, such as an amplitude mismatch of the rate of change  of the distribution function. In this paper we compare the theoretical predictions of Chandrasekhar's theory for anisotropic clusters to precise $N$-body data.  We show that more careful $N$-body measurements are able to reduce the claimed mismatch. Nevertheless, we observe a dependency of    the remaining mismatch on both position and anisotropy.   While the dependence on anisotropy may be understood qualitatively on theoretical grounds, the radial dependence implies that spatial inhomogeneities,  
and therefore collective effects, may become unavoidable to resolve the  residual mismatch between theory and simulations.

}

\keywords{
Diffusion -- Gravitation -- Galaxies: kinematics and dynamics
}

\maketitle

\section{Introduction}
\label{sec:intro}

 Two-body relaxation is one of the fundamental processes in stellar dynamics, of particular relevance to the dynamics of globular star clusters.  The most enduring theory is due to \citet{Chandrasekhar1941}, with extensions which will be mentioned later.  It depends on a number of approximations, including the assumption that the stellar density is uniform, and so the question of its accuracy naturally arises. No practical numerical methods existed at the time, and attempts to compare Chandrasekhar's theory with numerical simulations had to wait about 30 years.  An early example is \citet{Aarseth1974}, in which the evolution from Plummer initial conditions was followed up to about core collapse.  Chandrasekhar's theory was represented by a Monte Carlo model, which can be thought of as a solution of the orbit-averaged Fokker--Planck equation, based on Chandrasekhar's theory of relaxation. This equation, in its orbit-averaged form, is one of the extensions of the theory which we assume when we refer to ``Chandrasekhar's theory" in the present paper.\footnote{We shall also use the short notation ``NR", which stands for ``non-resonant", as opposed to the resonant theory of relaxation \citep[see, e.g.,][]{Heyvaerts2010}.}  The $N$-body models used up to 250 particles.

Another difference from the original form of the theory is in the ``Coulomb logarithm", which affects the rate at which two-body relaxation acts in the theory.  This depends on the maximum impact parameter of encounters included in the theory, which Chandrasekhar took to be the mean interparticle distance.  But \citet{Henon1958} argued that it should be of the order of the size of the system, which had the effect of increasing the rate of relaxation by a factor of 3/2. H\'enon's value became  universally accepted, but again we refer to the theory as Chandrasekhar's theory, even though the value of the Coulomb logarithm is H\'enon's. We return to this aspect of the theory in Section~\ref{sec:global-comparisons}.

What \citet{Aarseth1974} concluded was that Chandrasekhar's theory, as implemented in the Monte Carlo models, caused the system to evolve about 1.5 times faster than in the $N$-body simulations.  But in the present day, simulations routinely use much larger numbers of particles ($N \sim 10^5$) than those used in the 1974 paper ($N\le 250$).  Similarly, Monte Carlo models have improved greatly, thanks in part to comparisons with $N$-body data \citep[see, for example, section~3 of ][]{Giersz1998}.  

 This last point implies that we should avoid using Monte Carlo models, as \citet{Aarseth1974} did, to assess the difference between Chandrasekhar's theory and $N$-body simulations.  Fortunately, there is an alternative, which is a numerical solution of the Fokker--Planck equation by \citet{Takahashi1995} using a finite element method.  It gives anisotropic solutions, and is the best available method in terms of energy conservation, for example.  In Section~\ref{sec:global-comparisons}  we use this and other modernizations to reexamine the kind of comparison carried out by \citet{Aarseth1974}.

 The comparison by  \citet{Aarseth1974}  is a {\sl global} one, comparing Chandrasekhar's theory with $N$-body data by means of a single number covering a significant section of the evolution (up to core collapse).  It is also possible to make a more fine-grained comparison.  Chandrasekhar's theory represents relaxation as a result of orbital diffusion, and it has been possible to test it by measuring the diffusion coefficients in $N$-body data, as in the work of \citet{Theuns1996}.   He studied the diffusion in energy as a function of the energy of the particle, concluding that the overall agreement between the $N$-body data and the theory was ``impressive", but could differ by a factor up to about 2 for both strongly bound and weakly bound particles, when the $N$-body results implied {\sl stronger} diffusion than theory. 

Checking diffusion coefficients is perhaps a rather weak way of testing the validity of Chandrasekhar's relaxation theory, for the following reason.  The denser central parts of the $N$-body system are nearly isothermal (or quickly become so).  It is for this reason that the time scale of evolution is several relaxation times (cf. Section~\ref{sec:global-comparisons}), and the effect is that, in the collision term of the Fokker--Planck equation, there is a near-balance between the terms in the first and second moments.  Therefore, the diffusion coefficients in Chandrasekhar's theory might be good to 10\%, say, but the  near-balance  between the terms might lead to an overall error of order 1 or more in the collision term as a whole.  As such, a more sensitive test would be to compare the collisional flux in phase space, or even the rate of change of the distribution function itself (which is the divergence of the flux).

 The comparison of the flux was pioneered by \citet{Lau2019,Lau2021}\footnote{The two papers need to be read together, as the second is an important erratum.}, who computed the flux in the phase space of radial action and angular momentum.  The conclusion of the second paper was that the result from Chandrasekhar's theory ``quite closely" resembled that from $N$-body simulations, except for nearly radial orbits. However, no quantitative estimate of the agreement was attempted.  

Comparison of $dF/dt$ -- the rate of change of the distribution function -- was finally achieved by \citet{Tep2022} (see, especially,  their section~3.3), which we shall refer to as \citetalias{Tep2022}.  They studied the early evolution of the isotropic Plummer model (and some of its anisotropic variants) with $N = 10^5$ particles and 100 realizations, though fewer for some variants.  By averaging the results across different regions of phase space, the authors summarized the data by concluding that the theory overestimated the rate of diffusion by a factor of about 1.4 for the classical Plummer model, and that the discrepancy was greatest for the region inside the half-mass radius.  

 Although this conclusion is quite similar to that of  \citet{Aarseth1974}, these two results differ in sign from that of  \citet{Theuns1996}.  Evidently some reconciliation is needed, and this is what is attempted in the present paper. While this might seem an academic exercise, it is necessary, because the present incarnations of the Monte Carlo method \citep{Giersz1998,Joshi2000} have at their foundation H\'enon's version of Chandrasekhar's theory,  and these codes are themselves a basic tool for the study of the dynamical evolution of rich star clusters.

In the next sections, we present updated treatments of two of the results presented above: Section~\ref{sec:theory} first presents a single global number for the match between theory and simulation over the entire evolution from a Plummer model to core collapse, in analogy with \citet{Aarseth1974}, then presents a new analysis of  a much larger dataset of the kind presented by \citetalias{Tep2022}. Section~\ref{sec:conclusions} presents some discussion intended to illuminate why the revisions are needed, and summarizes the conclusions.

\section{Secular relaxation theory of spherical systems}
\label{sec:theory}

\subsection{Global comparisons}\label{sec:global-comparisons}

In this section, our aim is to update the study of \citet{Aarseth1974}, which compared the results of $N$-body simulations with a number of Monte Carlo solutions of the Fokker--Planck equation, on a time interval covering most of the time to core collapse, which we denote by $t_{cc}$.  There are two main improvements that we adopt: (i) on the $N$-body simulations  and the inferred value of $dF/dt$; and (ii) on the numerical evaluation of $dF/dt$ given by  the Fokker--Planck equation.

For the $N$-body simulations, we adopt the results of  \citet{Pavlik2018}, who computed the evolution of 85 models with $N=10^4$ particles and 10 models with $N=5\times10^4$ particles, all models starting from the Plummer model with equal masses.  Their results, given in Tables 1 and A.2 in the cited paper, lead to the conclusion that the time to core collapse is 
\begin{equation}\label{eq:tccNB}
    t_{\rc \rc,\mathrm{NB}} = \begin{cases}
            (2297 \pm 52) \,\rHU,& \text{for $N= 10^4$},\\
            (9347 \pm 150) \,\rHU,& \text{for $N = 5\times 10^4$},
\end{cases}
\end{equation} 
where the error bounds are 1$\sigma$ in the standard error and HU means H\'enon units.\footnote{Actually Pavlik and \u Subr give two values for the suite of larger systems, because of uncertainty (in some cases) about which of two deep collapses is to be regarded as ``the" core collapse.  The main text gives the result for the first core collapse, and the result for the second was $t_{cc,NB} = 9575 \pm 118$ HU.}

 For the numerical solution of the Fokker--Planck equation, we avoid the use of the Monte Carlo method, for the reason mentioned in Section~\ref{sec:intro}, viz., that $N$-body data may have been used in the selection of parameters of the Fokker--Planck code.  Non-stochastic Fokker--Planck solvers have been used for the study of core collapse back to \citet{Cohn1979,Cohn1980}.  Both are suboptimal, because the anisotropic treatment of the earlier paper led to poor energy conservation, while the later paper adopted an isotropic model.  These were finite-difference solutions, and later it was found that finite-element methods led to much better energy conservation in the anisotropic case \citep{Takahashi1995}.  Therefore we adopt these results, in which the time to core collapse is given as 
\begin{equation}
    t_{\rc\rc,\mathrm{FP}} = 17.6\, \trh,\label{tccFP}
\end{equation} 
where  $\trh$ is the initial half-mass relaxation time.  For the Plummer model with $N$ particles, this can be expressed in H\'enon units as 
\begin{equation}
    \trh = 0.093\frac{N}{\ln\Lambda},\label{eq:trhi}
\end{equation}  
where $\ln\Lambda$ is the Coulomb logarithm.

The final ingredient needed for a comparison between $N$-body and Fokker--Planck results is the value to be adopted for the argument of the Coulomb logarithm.  Commonly used is the value $\Lambda = 0.11N$ obtained by \citet{Giersz1994}. However, this depends on $N$-body results and thus should be avoided for the same reason that we avoid Monte Carlo methods. Another commonly used value is $\Lambda = 0.4N$ \citep{Spitzer1969}, which was the value adopted by \citet{Aarseth1974}. But the derivation of this formula ignores so-called ``non-dominant" terms in the calculations, i.e., those which are negligible in comparison with $\Lambda$ itself for large $N$, albeit only logarithmically\footnote{The term itself was introduced in this context by section~2.2(iii) of \citet{Chandrasekhar1942}.}. \citet{Henon1975} repeated the calculation while including such terms, obtaining the value
\begin{equation}
\Lambda = 0.15N,   \label{eq:Lambda}  
\end{equation}
which we shall use henceforth.

At this point, the job is almost done.  Combining equations~\eqref{eq:tccNB}--\eqref{eq:Lambda}, we immediately find that 
\begin{equation}
    \frac{t_{\rc\rc,\mathrm{NB}}}{t_{\rc\rc,\mathrm{FP}}} =
    \begin{cases}
        1.026 \pm 0.023 & \text{for $N = 10^4$},\\
        1.019 \pm 0.016 & \text{for $N=5\times10^4$}.
    \end{cases}
\end{equation}
By these measures, Chandrasekhar's theory and $N$-body simulations for the isotropic Plummer cluster agree to within a few percent.

\subsection{Theoretical prediction of $\partial F/\partial t$}\label{sec:dfdt-comparisons}

We use the orbit-averaged Chandrasekhar theory (see, e.g., \citetalias{Tep2022}, for a study of anisotropic clusters) to make theoretical predictions of the relaxation rate, $\p F/\p t$, and compare to ensemble-averaged measures made in $N$-body simulations. This theory predicts that the  evolution of the DF in action space is driven by  an orbit-averaged Fokker--Planck equation in action space  \citep[see, e.g., \S{7.4} of][]{Binney2008}
\begin{align}
\label{eq:def_FP}
\frac{\p F (\bJ,t)}{\p t} {} &\!=\! - \frac{\p }{\p \bJ} \!\cdot\! \mbF (\bJ) \\
&\!=\! - \frac{\p }{\p \bJ}\! \cdot \!\bigg[  \bD_{1} (\bJ) \, F (\bJ)\!-\! \frac{1}{2} \frac{\p }{\p \bJ} \!\cdot\! \bigg(  \bD_{2} (\bJ) \, F (\bJ)  \bigg) \bigg] ,\notag
\end{align}
with $\mbF (\bJ)$ the action space flux and $F=2L \Ftot$ the reduced DF in $(\Jr,L)$ space. The diffusion coefficients can be explicitly computed from the local velocity deflection coefficients. 
We refer to Appendix~\ref{app:NR_theory} for a summary of Chandrasekhar's theory. 

Once $\p F/\p t$ has been computed, we have access to the initial time evolution of a variety of dynamical quantities, such as that of the potential $\p \psi/\p t$ and of the core radius $\rd \Rc/\rd t$. We show in Appendix~\ref{app:potential_Rc} that these quantities are linearly related to $\p F/\p t$, and can be evaluated by using the matrix method.

In this paper, we shall study a family of anisotropic Plummer spheres \citep{Dejonghe1987} subject to the Plummer potential
\begin{equation}
\psi (r) = - \frac{G M}{\sqrt{b^{2} + r^{2}}},
\label{Plummer_potential}
\end{equation}
where $b$ is the Plummer scale length, 
and parameterized by $q \in (-\infty,1)$ 
such that the (velocity) anisotropy parameter reads
\begin{equation}
\beta(r) = 1-\frac{\sigma^2_{\rt}}{2\sigma^2_{\rr}} = \frac{q}{2} \frac{r^2}{b^2+r^2}.
\end{equation}

\subsection{Measure of the relaxation rate in $N$-body runs}

In order to measure the relaxation rate, we must compute a time derivative from the $N$-body data. To do so, we follow \citetalias{Tep2022} and employ at first a finite  difference  scheme
\begin{equation}
\label{eq:FD}
G(\bJ,T) = \frac{F(\bJ,T)-F(\bJ,0)}{T},
\end{equation}
where the two endpoints are chosen to be the initial time, $0$, and some final time $T$. This endpoint $T$ must be chosen appropriately, so that
$G(\bJ,T)$ is a good estimator of $\p F/\p t$   at $t=0$. In addition, due to the inherent discrete nature of our sampling, we have to estimate the DF $F(\bJ)$ on a homogeneous grid of actions $\bJ_i$. This is done by binning action space, as described in Appendix~\ref{app:Nbody_measures}. 

Already, we can see that the measure of relaxation rate, $\p F/\p t$, in $N$-body runs is impacted by three parameters: the final time $T$, the size of radial action bins $\delta \Jr$ and the size of angular momentum bins $\delta L$.
Choosing these parameters is not a trivial task. Indeed, we must choose $T$ to be small enough so that the finite   difference scheme yields a   satisfactory approximation, but we must   not  pick too small a value so as not to be dominated by finite-$N$ fluctuations. A similar balance occurs for the choice of bin size, where we must select a bin small enough not to smooth out any signal, but not too small so as not to be dominated by fluctuations.

\begin{figure*} 
    
    \begin{minipage}[h]{1.0\linewidth}
    \centering
   \includegraphics[width=0.85\textwidth]{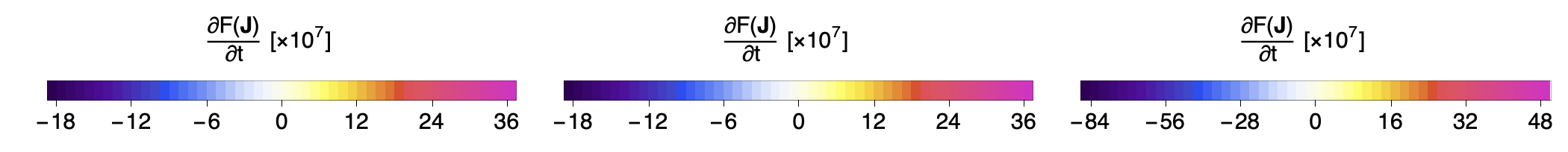}
    \end{minipage}

    \vspace{-2.5mm}

     \begin{minipage}[h]{1.0\linewidth}
    \centering
    \includegraphics[width=0.85\textwidth]{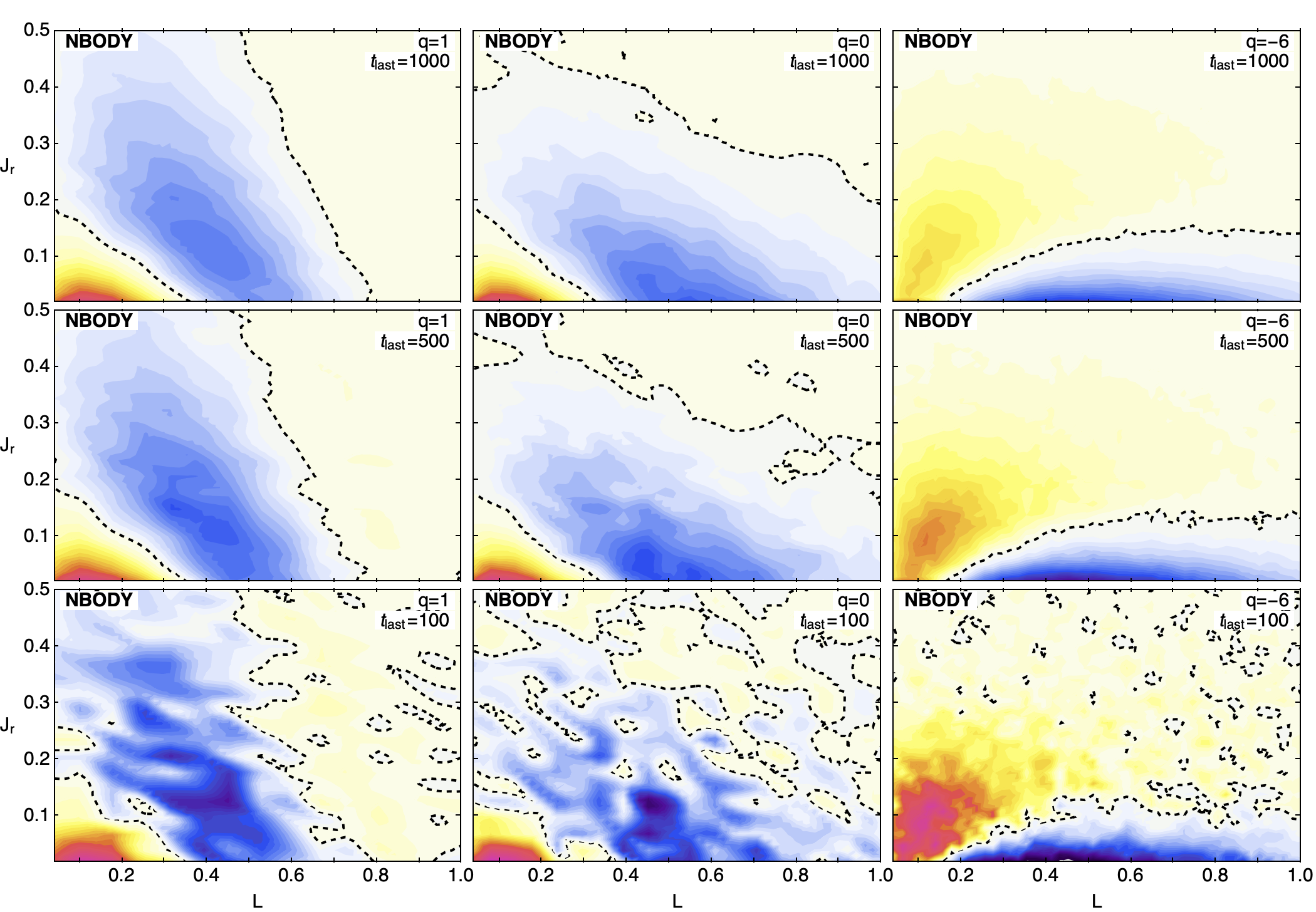}
   \end{minipage}
   
   \caption{Illustration of the impact of the choice of $T$ on the relaxation rate, $\p F/\p t$ -- ensemble-averaged over 100 realizations -- for three types of anisotropic clusters, going from radially (leftmost panels) to tangentially anisotropic clusters (rightmost panels). Decreasing the final time $T$ (from top panels to bottom panels) beyond a certain point destroys the smooth structures due to fluctuations, making the finite difference calculation no longer a correct estimator of the time derivative of the mean-field DF. 
      }
   \label{fig:dFdt_tlast}
 \end{figure*}

Following \citet{Feliachi2024}, we can show that the fluctuations around the ensemble-averaged relaxation rate follow the limit behavior
\begin{equation}
\label{eq:large_deviation}
\sigma [G ] (\bJ) \sim \frac{1}{\delta \Jr \delta L\sqrt{T }},
\end{equation}
which diverges for  small choices of parameters. We illustrate this effect in Figure~\ref{fig:dFdt_tlast}, where we represent the relaxation for various values of $T$.

Previous works made the choice of picking some  intermediate value $T$ and $\delta \Jr \delta L$, such that the qualitative structure of the relaxation rate has converged while remaining smooth enough. However, choosing an intermediate value for these parameters impacts the amplitude of the result.
\begin{figure} 
    \centering
   \includegraphics[width=0.45\textwidth]{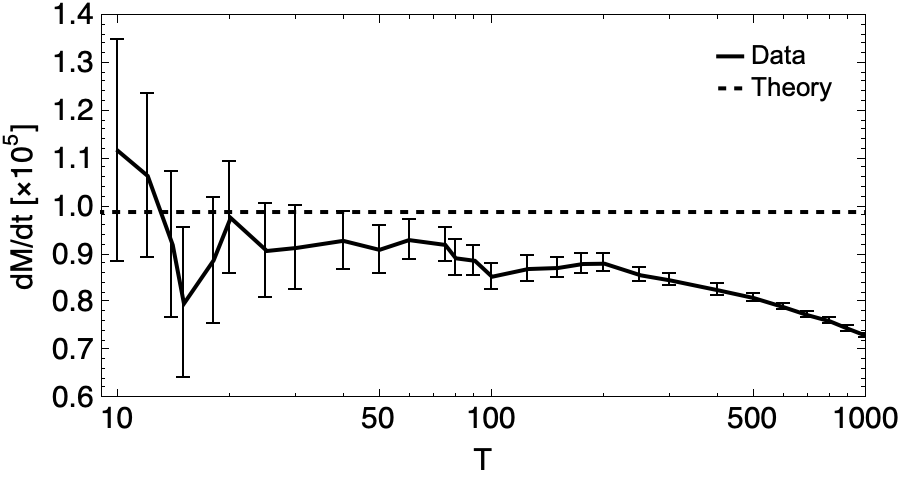}
   \caption{Evolution of the central enclosed mass, $\rd M/\rd t$, obtained from $N$-body measurements w.r.t. the final time of the finite difference scheme, $T$, for the isotropic cluster. The enclosed region is defined by the positivity of $\p F/\p t$, as measured in Figure~\ref{fig:dFdt_interp} (middle panel). We show the Chandrasekhar theoretical prediction as a dashed line. As one considers small $T$, the measurement gets closer to the theoretical prediction -- typically going from a $\sim\! 40 \%$ error at ${T\!=\!1000\,\rHU}$ to a $\sim\! 10\%$ error at ${T\!=\!100 \, \rHU}$. However, at the same time, the fluctuations make it difficult to obtain a precise measurement.  }
   \label{fig:dMdt_T}
 \end{figure}
Figure~\ref{fig:dMdt_T}\footnote{ This figure shows results for a measure of $F$ which is different from $F$ itself (see the caption). But in terms of the foregoing discussion, it exhibits the same behaviour as does the evolution of $F$.} shows that the error in the  rate  of relaxation can reach up  to about 40\% of the extrapolated $T \rightarrow 0$ limit we  infer. The error bars we obtained from measurements diverge as $T \rightarrow 0$, as predicted by  equation~\eqref{eq:large_deviation}.    It is clear that the estimation of $\p F/\p t$ as $t\to0$ requires a better method.

\subsection{Comparison between Chandrasekhar predictions and $N$-body measurements}{\dchchange \label{comparison}}

\begin{figure} 
    \centering
   \includegraphics[width=0.45\textwidth]{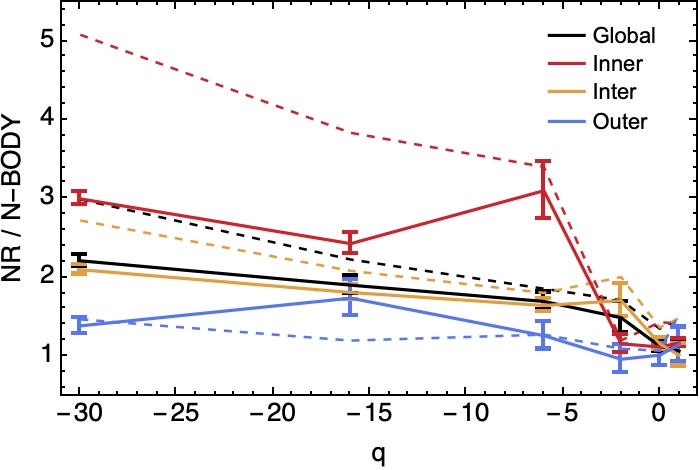}
   \caption{Ratio NR/$N$-body, as defined by equation~\eqref{eq:Ratio_NR_NBODY}. The error bars correspond to a  $1\sigma$ confidence interval, computed from all realizations for a given anisotropy. We compare the results from this paper's fitting method (full lines, see Table~\ref{table:ratioNRNBODY_table}) to those of \citetalias{Tep2022} (dashed lines). Applying a finer analysis of the $N$-body data through the linear fit and least-squares regression reduces the amplitude mismatch with Chandrasekhar's theory. Nevertheless, we cannot fully get rid of this mismatch altogether, and still observe a slightly increasing mismatch as we consider more tangentially anisotropic clusters. }
   \label{fig:ratio_NR_NBODY}
 \end{figure}

To estimate that limit, we shall now fit the early time evolution of the DF at each action bin. First, we can use a polynomial fit of the DF  (as a function of time) at each bin,  setting the degree as a free parameter which we must choose. We detail the procedure in  Appendix~\ref{app:poly_fit}. By fitting the time evolution of the DF over the first 1000 HUs by a polynomial expansion, we can compute an estimate of $\p F/\p t$, which is both   nearly  independent of the choice of $T$ and closer to the real ensemble average measurement (see Figure~\ref{fig:interpolation_DF}). 
 Then we show in Appendix~\ref{app:interpolation} (see Figure~\ref{fig:interpolation_DF} and \ref{fig:interpolation_LR_DF})  that the DF follows a linear behavior during the first 100 HUs. Therefore, we can use a linear model (i.e. a polynomial expansion with degree 1) to fit the DF time evolution without impacting the estimation of $\p F/\p t$, provided we restrict the fitting interval to the first 100 HUs.   This is the method we use in all the substantive results presented subsequently in the main text of this paper (except in Figure~\ref{fig:ratio_NR_NBODY}, which compares results of this method to those obtained by finite differences (Equation~\ref{eq:FD})).

\begin{figure*} 

    \begin{minipage}[h]{1.0\linewidth}
    \centering
   \includegraphics[width=0.75\textwidth]{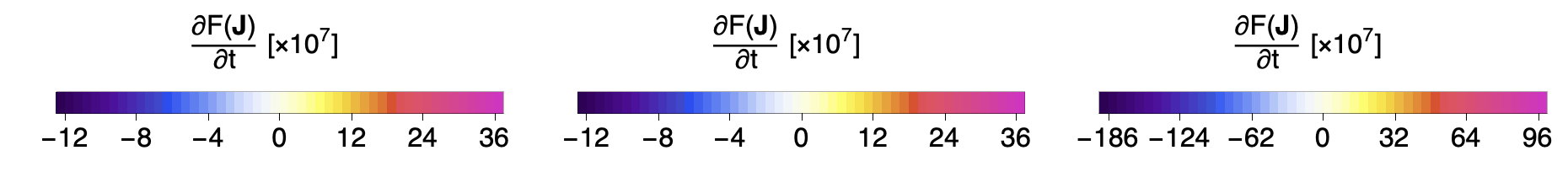}
   \end{minipage}

 \vspace{-2.5mm}

    \begin{minipage}[h]{1.0\linewidth}
    \centering
\includegraphics[width=0.80\textwidth]{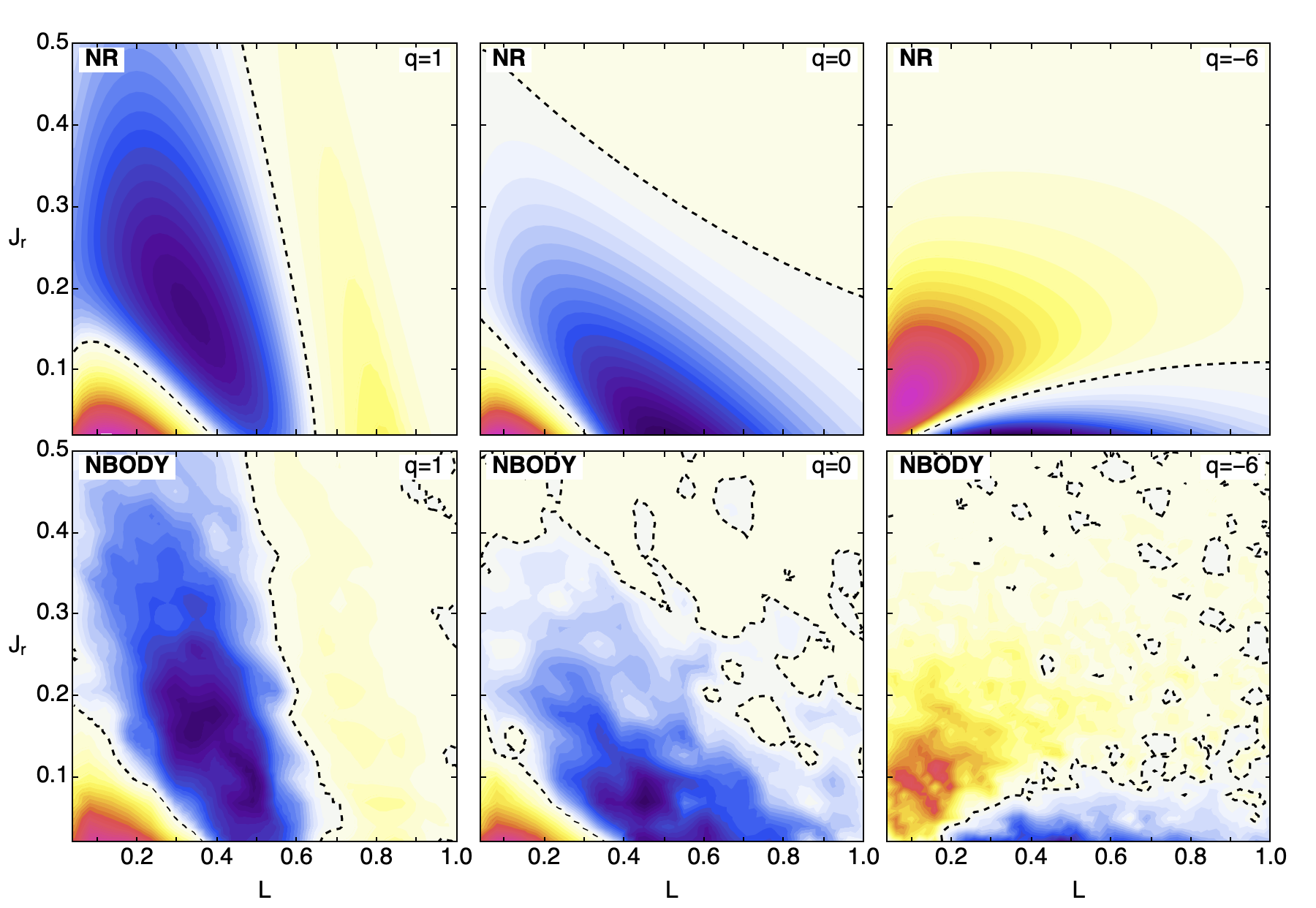}
    \end{minipage}
   \caption{Illustration of the relaxation rate, $\p F/\p t$, for the Chandrasekhar prediction (top panels) and $N$-body measurements (bottom panels) and a variety of anisotropies. The $N$-body measurement is obtained from the linear interpolation scheme described in Appendix~\ref{app:poly_fit}. The $N$-body measurements matches the structures (boundary between positive and negative values of $\p F/\p t$, qualitative behavior of its amplitude) in action space predicted by theory}, and appear smoother compared to the finite difference estimations for low $t_{\rm last}$ shown in Figure~\ref{fig:dFdt_tlast}, while their overall amplitudes are closer to the Chandrasekhar prediction. 
      
   \label{fig:dFdt_interp}
 \end{figure*}

We show a few  maps  of these results in  Figure~\ref{fig:dFdt_interp}.
Of course, we recover in both theoretical predictions and $N$-body measurements the usual isotropization process of anisotropic cluster, as well as the impact of core collapse in the central regions of the cluster (bottom left region of action space) and of star evaporation in the outer regions (top right region of action space). We refer to figure~{F1} of \citetalias{Tep2022} for more details. Furthermore,  the amplitudes we measure in $N$-body simulations through this method are much closer to the Chandrasekhar predictions than the ones we obtained in Figure~\ref{fig:dFdt_tlast}. 
Indeed, the measurements shown in Figure~\ref{fig:dFdt_interp}  are obtained through linear fitting of the (ensemble averaged) $N$-body data, and they estimate the value of $\partial F/\partial t$ at $t=0$. As such, the way they should be compared to the panels of Figure~\ref{fig:dFdt_tlast}  is by extrapolating the behavior of the latter ones as $t_{\rm last}\rightarrow 0$. In that limit, the finite difference scheme only becomes more and more fluctuating, which is not the case of the linear fitting scheme. In the case of $q=-6$, the finite difference scheme appears satisfactory at first because most particles are located near the $J_r=0$ axis and contributed to the relevant bins. This, in effect, reduces the fluctuations, at least for values of $t_{\rm last}$ that are not too small.

To quantify this observation, we define the ratios $\text{NR}/N\text{-BODY}$ 
\begin{equation}
\frac{\text{NR}}{N\text{-BODY}} \bigg|_F= \frac{\displaystyle{\int \rd \bJ \, F(\bJ)\, |\p F / \p t|_{\text{NR}}}}{\displaystyle{\int \rd \bJ \, F(\bJ)\, |\p F / \p t|_{N\text{-BODY}}}},
\label{eq:Ratio_NR_NBODY}
\end{equation}
as well as an alternative quantity
\begin{equation}
\frac{\text{NR}}{N\text{-BODY}} = \frac{\displaystyle{\int \rd \bJ \,  |\p F / \p t|_{\text{NR}}}}{\displaystyle{\int \rd \bJ \, |\p F / \p t|_{N\text{-BODY}}}},
\label{eq:Ratio_NR_NBODY_alt}
\end{equation}
where the integration weight is uniform. 
Then, we can use the estimated values of $\p F/\p t$ to estimate the ratios given by equations~\eqref{eq:Ratio_NR_NBODY} and \eqref{eq:Ratio_NR_NBODY_alt}. We gather the results in Table~\ref{table:ratioNRNBODY_table} and Table~\ref{table:ratioNRNBODY_table_alt},
\begin{table}[h!]
\centering
\setlength{\tabcolsep}{.3em}
 \begin{tabular}{|c|c|c|c|c|c|}
\hline
$q$ & Global &Inner & Mid &  Outer & $t_{\mathrm{end}}$ \\
\hline
\hline
1   & $1.06 \pm 0.14$   &  $1.17 \pm 0.05$ & $1.01 \pm 0.14$  &   $1.15 \pm 0.22$ & 100 \\
0   & $1.13 \pm 0.07$   &  $1.12 \pm 0.02$ & $1.16 \pm 0.07$  &   $1.01 \pm 0.13$ & 100 \\
-2  & $1.49 \pm 0.20$   &  $1.16 \pm 0.11$ & $1.70 \pm 0.21$  &   $0.96 \pm 0.18$ & 100 \\
-6  & $1.69 \pm 0.11$   &  $3.10 \pm 0.36$ & $1.64 \pm 0.08$  &   $1.26 \pm 0.18$ & 50\\
-16 & $1.90 \pm 0.11$   &  $2.42 \pm 0.13$ & $1.80 \pm 0.09$  &   $1.74 \pm 0.23$ & 50 \\
-30 & $2.21 \pm 0.07$   &  $3.00 \pm 0.08$ & $2.10 \pm 0.06$  &   $1.38 \pm 0.10$ & 50 \\
\hline
\end{tabular}
\caption{Ratio NR/$N$-body evaluated using equation~\eqref{eq:Ratio_NR_NBODY}. We evaluate the $N$-body contribution using the linear fitting scheme described previously. $t_{end}$ is chosen to be lower for higher tangential anisotropies because relaxation occurs on a shorter timescale  (see Table \ref{table:dRcdt_table} below). Whereas isotropic clusters have a global ratio close to $1$, this value increases as one considers more  tangentially anisotropic clusters. This behavior seems to occur in each region of the clusters, though the quantity ratio itself appears to depend on the location in the cluster.} 
\label{table:ratioNRNBODY_table}
\end{table}
and compare those to previous measurements of \citetalias{Tep2022} in Figure~\ref{fig:ratio_NR_NBODY}. 
We observe that the mismatch between theory and prediction has been reduced by a large factor. This is especially striking for the isotropic cluster, where a proper estimation of the $\p F/\p t$ yields a global ratio of about 1. However, the fact remains that increasing initial tangential anisotropy increases the mismatch between theory and $N$-body measurements. 
This dependency can be partially understood by computing the effect of anisotropy on the Coulomb logarithm that is present inside the numerator of these ratios. We show in Appendix~\ref{app:Coulomb_parameter_anisotropy} that the Coulomb anisotropy tends to decrease as the system strays from isotropy. This reduces some of the remaining mismatch between theory and simulations, although it does not completely resolve it.
\begin{figure} 
    \centering
   \includegraphics[width=0.45\textwidth]{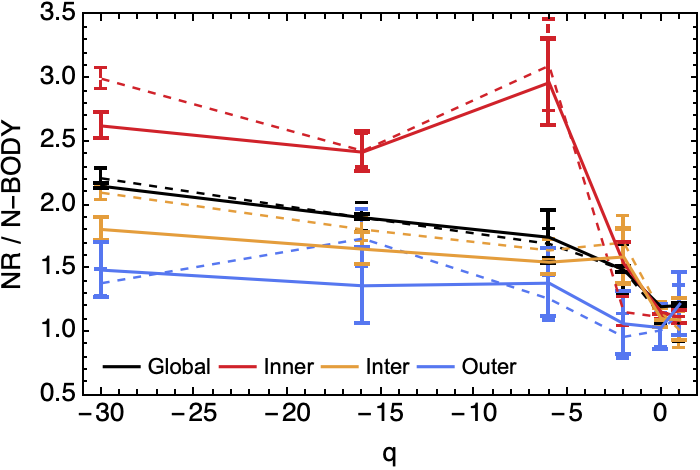}
   \caption{Comparison of the two NR/$N$-body ratios, defined by equation~\eqref{eq:Ratio_NR_NBODY} (DF-weighed, in dashed lines) and equation~\eqref{eq:Ratio_NR_NBODY_alt} (uniformly weighed, in full lines). The error bars are computed in the same way as in Figure~\ref{fig:ratio_NR_NBODY}.  We do not observe any significant qualitative differences in their behavior w.r.t. anisotropy, nor do we observe any strong quantitative differences.
      }
   \label{fig:ratio_NR_NBODY_compare}
 \end{figure}

\begin{table}[h!]
\centering
\setlength{\tabcolsep}{.3em}
 \begin{tabular}{|c|c|c|c|c|c|}
\hline
$q$ & Global &Inner & Mid &  Outer & $t_{\mathrm{end}}$\\
\hline
\hline
1   & $1.21 \pm 0.01$   &  $1.12 \pm 0.05$ & $1.10 \pm 0.17$  &   $1.22 \pm 0.25$ & 100 \\
0   & $1.20 \pm 0.01$   &  $1.16 \pm 0.02$ & $1.11 \pm 0.07$  &   $1.03 \pm 0.18$ & 100 \\
-2  & $1.49 \pm 0.02$   &  $1.54 \pm 0.17$ & $1.59 \pm 0.22$  &   $1.06 \pm 0.25$ & 100 \\
-6  & $1.75 \pm 0.02$   &  $2.97 \pm 0.34$ & $1.55 \pm 0.10$  &   $1.38 \pm 0.27$ & 50\\
-16 & $1.90 \pm 0.02$   &  $2.42 \pm 0.15$ & $1.65 \pm 0.12$  &   $1.36 \pm 0.30$ & 50 \\
-30 & $2.15 \pm 0.02$   &  $2.62 \pm 0.10$ & $1.81 \pm 0.09$  &   $1.49 \pm 0.22$ & 50 \\
\hline
\end{tabular}
\caption{Alternative ratio NR/$N$-body, evaluated using equation~\eqref{eq:Ratio_NR_NBODY_alt}. We observe the same trends as in Table~\ref{table:ratioNRNBODY_table}.
}
\label{table:ratioNRNBODY_table_alt}
\end{table}

 Indeed, the ratio also appears to depend on the location within the cluster, with a mismatch especially important in the central region of the cluster.  It has long been suggested \citep{Spitzer1987, HeggieHut2003} that  the Coulomb logarithm should take a smaller value at smaller radii. The reason for this is that its argument is proportional to the maximum impact parameter for encounters; this in turn is often taken as the half-mass radius, but in the core encounters will be suppressed at such a large radius because of the density profile, and it may be better to choose the core radius.  Thus the Coulomb logarithm will be smaller in the core than globally, depressing the values of NR used in Tables 1 and 2, and tending to improve the agreement with N-body data. 

 These arguments are qualitatively appealing, but quantitatively they seem insufficient.  The results in Tables 1 and 2 for the isotropic case ($q=0$) suggest that any variation in $\Lambda$ between small and large radii might be of order 10\%, which would be incapable of accounting for the variation in the mismatch between small and large radii for the highly tangentially anisotropic models in the Tables; these suggest a variation of order 100\%.  The dependence on anisotropy calculated in Appendix E also seems insufficient: while the value of $\gamma$ may vary by a factor of 3 in Figure~\ref{fig:gamma_j}, the resulting variation in $\Lambda = \gamma N$ is only of order 10\% when $N=10^5$, whereas the data in Tables 1 and 2 suggests it would need to be of order 100\%.   Thus it seems likely that 
collective effects, which are beyond the reach of Chandrasekhar's theory, may play a crucial role in completely resolving the observed mismatch.

\begin{figure} 
    \centering
\includegraphics[width=0.45\textwidth]{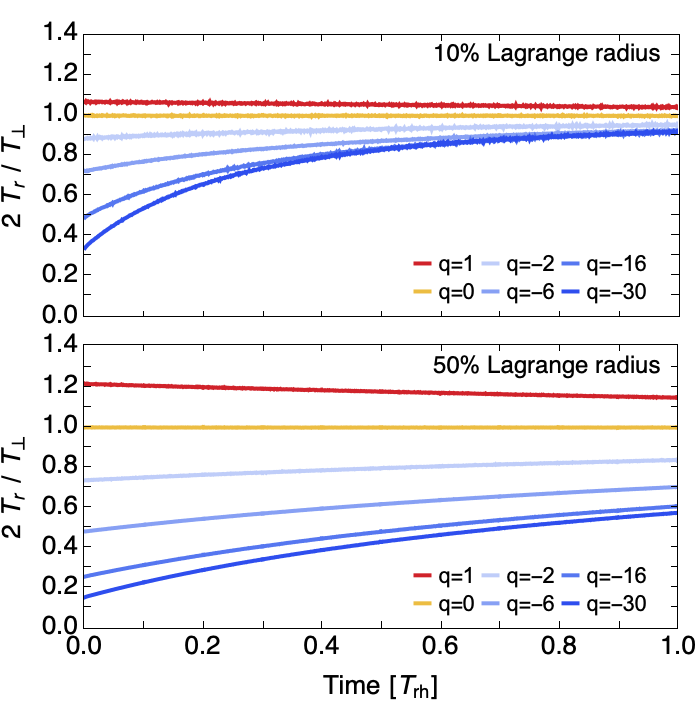}
   \caption{Early evolution of the anisotropy, $2\, T_{\mathrm{r}}/T_{\perp}$ \citep[see, e.g.,][for a definition]{Breen2017}, of a panel of clusters composed of $N=10^5$ stars. \textit{Top panel:} Evolution of the ratio within the 10 \% Lagrange radius. \textit{Bottom panel:} Evolution of the ratio the 50 \% Lagrange radius. We observe an overall isotropization of the cluster, which occurs faster as we consider its innermost regions. In particular, the region within the 10 \% Lagrange radius has become mostly isotropic within the first initial half-mass relaxation time. Runs have been ensemble averaged over 200 realizations for $q=0, -6$, over 100 realizations for $q=1$ and over 50 realizations $q=-2,-16,-30$. }
   \label{fig:isotropisation_Breen}
 \end{figure}

Another factor that should be taken into account is the time-dependence of anisotropy: the isotropization process undergone by anisotropic Plummer spheres happens on relatively quick timescales. Figure~\ref{fig:isotropisation_Breen} shows that the cluster is mostly isotropic by the first half-mass relaxation time within the 10\% Lagrange radius, and has already greatly converged towards that state within the 50\% Lagrange radius. As such, the impact of the larger ratios we obtained for anisotropic clusters might not play as important a role (e.g. in the time to core collapse) as one could believe at first.

\subsection{Estimating the mismatch from other quantities}

While the study of $\p F/\p t$ allows us to examine the cluster's relaxation in both a global picture (through global ratios and the global relaxation in action space) and a local picture (by probing separately the inner, intermediate and outer regions of the cluster) relatively straightforwardly, we showed that the measurement of this quantity was highly non-trivial and very sensitive to finite-$N$ effects. Therefore, we shall complement our previous analysis with the study of two additional quantities which are less affected by these issues.{\footnote{ See also the data on the "central enclosed mass" in Figure~\ref{fig:dMdt_T}}}

\begin{figure*} 
    \centering
  \includegraphics[width=0.95\textwidth]{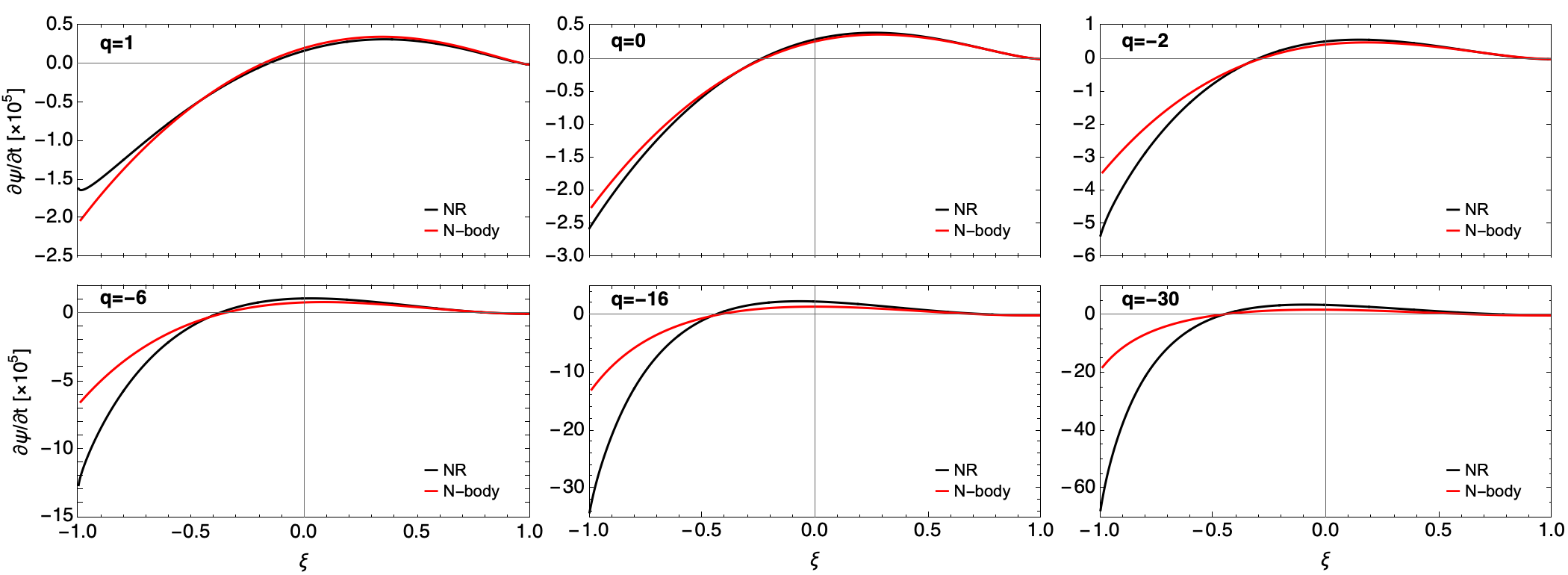}
   \caption{Evolution of the stellar potential, $\p \psi/\p t$, from $N$-body measurements (in red) and Chandrasekhar predictions (in black), for the ${q\!=\!1,0,-2}$ (top panel, left to right) and ${q\!=\!-6,-16,-30}$ (bottom panel, left to right)   clusters. $N$-body runs have been ensemble averaged over 100 realizations (resp. 50 realizations) for $q=1,0,-6$ (resp. $q=-2,-16,-30$), and plotted against the reduced radial variable $\xi=(r^2-b^2)/(r^2+b^2)$. The Chandrasekhar prediction of this rate of change closely matches that of simulations: it predicts the point  at which $
\p \psi/\p t$ vanishes and  for the most part only differs from the $N$-body measurements by an overall prefactor. }
   \label{fig:dpsidt}
 \end{figure*}

First, let us consider the cluster's potential, $\psi(r)$, which we will assume keeps its spherical symmetry during its evolution \citep[see, e.g., appendix~{F} of][]{Tep2024}. Its time derivative, $\p \psi / \p t$, is related to that of the density of the cluster, $\p \rho/\p t$, through a linear equation of the form $\p_t \rho = \mathcal{L}[\p_t \psi] + \mathcal{S}$. We can formally invert it and obtain the inverse relation $\p_t \psi =\mathcal{L}^{-1}[\p_t \rho - \mathcal{S}]$. In practice, this inversion requires the use of bi-orthogonal basis elements and the use of the matrix method. We detail this calculation in Appendix~\ref{app:potential_Rc}. 
The potential measurement in $N$-body simulation is quite straightforward,  and can be obtained using Hénon's method \citep{Henon1971} 
\begin{align}
\begin{array}{ll}
\begin{cases}
\psi_{N+1} &= 0 , \\ 
M_N &= M  ,
\end{cases} 
&\hspace*{3mm}
\begin{cases}
\psi_k &= \displaystyle{\psi_{k+1} - G M_k \big(r_k^{-1}- r_{k+1}^{-1} \big)} , \\
M_{k-1} &= M_k - m_k,
\end{cases} 
\end{array}
\end{align}
where $\psi_k$ is the potential at $r_k$ and $M_k\!=\!M(\leq r_k)$ is the mass within the sphere of radius $r_k$.
The time derivative is then estimated by finite differences. However, because the function depends only on the radial variable, this measurement is much less subject to fluctuations than that of $\p F/\p t$. 
We show in Figure~\ref{fig:dpsidt} the time derivatives of the potential for a selection of anisotropic clusters. 
The theory reproduces accurately the shape of the potential evolution, including the radius where $\p_t \psi=0$. We also observe   (at least at smaller radii) that the Chandrasekhar calculation overestimates the $N$-body measurement by a prefactor which increases with tangential anisotropy. 

Let us in turn consider the core radius, $\Rc$, defined by
\begin{equation}
\Rc^2 = \frac{\displaystyle{\int \rd \br \rho(\br)^3 r^2}}{\displaystyle{\int \rd \br \rho(\br)^3}}.
\end{equation}
Its time derivative is given by
\begin{equation}
 \dot{\Rc} = \frac{\displaystyle{3 \int \rd \br \dot{\rho}(\br) \rho(\br)^2 r^2}}{\displaystyle{2\Rc \int \rd \br \rho(\br)^3}} -  \frac{ \displaystyle{3 \Rc \int \rd \br \dot{\rho}(\br) \rho(\br)^2} }{\displaystyle{2\int \rd \br \rho(\br)^3}},
\end{equation}
and can be computed in a similar manner as for the potential (see  Appendix~\ref{app:potential_Rc} for details). We present a selection of values   in Table~\ref{table:dRcdt_table} for various initial cluster anisotropies. 
\begin{table}[h!]
\centering
\setlength{\tabcolsep}{.3em}
 \begin{tabular}{|c|c|c|c|c|}
\hline
$q$ & NR [$\times 10^5$] & $N$-body  [$\times 10^5$]& NR/$N$-body   \\
\hline
\hline
1 & -3.06& $-3.59 \pm 0.11$ & $0.85 \pm 0.03$\\
0& -5.00 & $-4.31 \pm 0.04$  & $1.16 \pm 0.01$ \\
-2 & -10.59 &$-7.31 \pm 0.12$ &$1.45 \pm 0.03$\\
-6 &-25.17 &$-16.14 \pm 0.17$ &$1.56 \pm 0.02$\\
-16 & -65.38  &$-38.41 \pm 0.33$ & $1.70 \pm 0.02$ \\
-30& -118.01 &$-64.31 \pm 0.55$& $1.84 \pm 0.02$\\
\hline
\end{tabular}
\caption{Relaxation rate, $\rd \Rc/\rd t$, as a function of cluster initial anisotropy, for both the Chandrasekhar theoretical prediction and $N$-body measurements. Increasing tangential anisotropy increases the rate of change of the core radius, which shows the acceleration of core collapse (\citealt{Breen2017}; \citetalias{Tep2022}). The rightmost column gives a ratio, analogous to equations~\eqref{eq:Ratio_NR_NBODY} and \eqref{eq:Ratio_NR_NBODY_alt}, and displaying the same trends as in Tables~\ref{table:ratioNRNBODY_table} and \ref{table:ratioNRNBODY_table_alt}. 
}
\label{table:dRcdt_table}
\end{table}
The $q=0$ value yields a ratio $\mathrm{NR}/N\mathrm{-body}$ of $1.16$, and greater tangential anisotropies yields increasing mismatches of the same order as the global ratios obtained in Table~\ref{table:ratioNRNBODY_table}.

The $q=1$ value for the rate of change of $\Rc$ has a different behavior compared to the others, which is related to the fact that the $N$-body measurement show a faster relaxation than the Chandrasekhar prediction.

\section{Conclusions and perspectives}
\label{sec:conclusions}

\subsection{Conclusion}

We re-investigated the effectiveness of orbit-averaged Chandrasekhar theory against a set of $N$-body simulations. We made a better estimation of the rate of change of the cluster's distribution function than in \citetalias{Tep2022} by carefully performing least squares regressions on the early time evolution of the cluster's DF. 

We applied these fitting methods to Plummer clusters with varying degrees of anisotropy, and were able to reduce the amplitude mismatch between prediction from Chandrasekhar theory and $N$-body measurement considerably, notably reaching a global ratio close to 1 in the isotropic case. However, this finer measurement method was not enough to completely erase any amplitude discrepancy between theory and simulations. Indeed, we can highlight two principal features of the mismatch. First, the mismatch depends on the position in the cluster, which goes against the constant Coulomb logarithm   usually adopted in the theory, though it is qualitatively consistent with some suggestions in the older literature. Second, the mismatch increases with initial tangential anisotropy. In particular, by revisiting work made by \citet{Henon1975}, we showed that this behavior may be due in part to a reduction of the Coulomb logarithm's value due to anisotropy. However,  these two effects alone might not be sufficient to completely resolve the remaining mismatch. Finally, we made similar qualitative and quantitative observations by considering other dynamical quantities, such as the rate of change of the cluster's potential and that of its core radius. Nevertheless, because initially anisotropic clusters appear to isotropize quickly within a few half-mass relaxation times, the impact of this increased prefactor mismatch might not be as important to the overall Chandrasekhar prediction of the cluster's relaxation as one might have believed at first.

\subsection{Perspectives}

  We have argued
that the observed mismatch between $N$-body data and Chandrasekhar theory, though qualitatively consistent with effects of anisotropy and inhomogeneity,  seems too large to be explicable quantitatively  by these effects, even though we  have not developed a systematic theory which includes them.
However, we do observe that the theoretical prediction and the numerical measurement of the  rate of change of the action space DF display the same structures.  Therefore,  a systematic theory accounting for these  effects  should be close to the Chandrasekhar one, in some sense. As such, one could seek to extend the work of \citet{Fouvry2021} to the case of anisotropic clusters, and compute the rate of change predicted by the inhomogeneous Landau theory, both of which reduce to the orbit-averaged Chandrasekhar theory in the homogeneous,  isotropic   limit \citep[see, e.g.,][and references within]{Tep2023}. 
Furthermore, since increasing tangential anisotropy increases the number of  near-circular orbits, it could be of interest to estimate the impact of coherent orbital interactions on the theoretical prediction. To that aim, the inclusion of the effect of  collective effects on the cluster's relaxation through the Balescu--Lenard equation \citep[see, e.g.][]{Heyvaerts2010} would be of interest.

In this paper, we used various least square fitting techniques to estimate the initial rate of change of the ensemble-averaged DF. This allowed us to smooth out fluctuations and somewhat reduce the number of necessary realizations. This came at the cost of introducing additional parameters such as the degree of the fitting polynomial or a time cutoff, which we had to fix by hand. It could be of interest to let go of this fitting method in favor of other fitting techniques -- such as total variation gradient \citep[see, e.g.,][]{Agarwal2003,Chartrand2011}. This should be coupled with the creation of  much larger number of cluster's realizations, in order to obtain a larger statistical sample.   And, as suggested by the referee, it may well be advantageous to sample more frequently than once per HU.  With such improvements one  would then be able to explore a variety of open questions, such as a more accurate prediction of the initial rate of change of the DF, which one could complement with an estimation of its RMS.

\section*{Data Distribution}

The  data underlying this article 
is available through reasonable request to the author.

\section*{Acknowledgements}

This work is partially supported  by the National Science Foundation under Grant No. AST-2310362 to UNC-Chapel Hill, by NASA ATP Grant 80NSSC24K0687, as well as the grant ExaSKAle ANR-24-CE31-5182 and  \href{https://www.secular-evolution.org}{\emph{SEGAL}} ANR-19-CE31-0017
of the French Agence Nationale de la Recherche.
This work has made use of the Infinity Cluster hosted by Institut d'Astrophysique de Paris, partially funded by IDF-DIM-ORIGINES-2023-4-11. 
We thank St\'ephane Rouberol for the smooth running of the
Infinity cluster. Last but not least, we are very grateful to the referee for comments which we hope have really helped us to clarify the presentation.

\appendix
\counterwithin{figure}{section}
\renewcommand\thefigure{\thesection\arabic{figure}}

\counterwithin{table}{section}
\renewcommand\thetable{\thesection\arabic{table}}

\section{Orbit-averaged Chandrasekhar theory}
\label{app:NR_theory}

Let us consider a spherically symmetric self-gravitating globular cluster with $N$ stars of the same individual mass $m=M/N$, where $M$ is the total cluster's mass.

Consider the orbit of a given test star plunged in that system. Its motion can be decomposed into a mean-field motion, imposed by the mean field of the spherically symmetric cluster, which is in turn perturbed by the finite-$N$ noise induced by the graininess of the potential. As a result, the test star undergoes a slow, irreversible diffusion of its orbital parameters driven by a succession of pairwise encounters with the cluster field stars. This induces a long-term relaxation, the so-called Chandrasekhar relaxation.

\begin{figure} 
    \centering
   \includegraphics[width=0.3\textwidth]{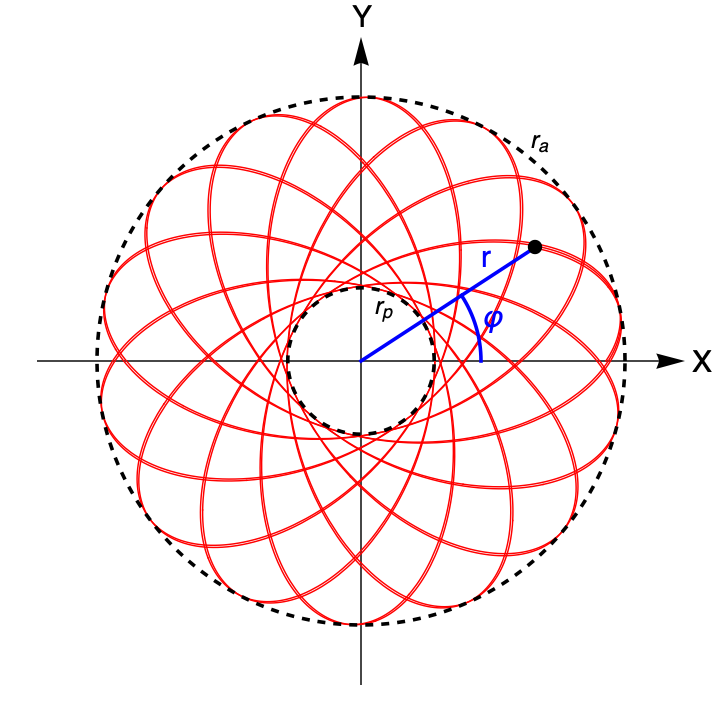}
   \caption{Figure 1 of \citet{Tep2024}. Illustration of the orbit average described by equation~\eqref{eq:generic_orbit_average}.
      }
   \label{fig:orbit_average}
 \end{figure}

Letting $\Ftot$ be the full  (anisotropic) DF of the cluster's stars, this process can be described by an orbit-averaged Fokker--Planck equation in action space  \citep[see, e.g., \S{7.4} of][]{Binney2008}
\begin{align}
\label{eq:def_FP}
\frac{\p F (\bJ,t)}{\p t} {} &\!=\! - \frac{\p }{\p \bJ} \!\cdot\! \mbF (\bJ) \\
&\!=\! - \frac{\p }{\p \bJ}\! \cdot \!\bigg[  \bD_{1} (\bJ) \, F (\bJ)\!-\! \frac{1}{2} \frac{\p }{\p \bJ} \!\cdot\! \bigg(  \bD_{2} (\bJ) \, F (\bJ)  \bigg) \bigg] ,\notag
\end{align}
where $\mbF (\bJ)$ is the action space flux and ${F\!=\!\int_{-L}^{L} \rd \Lz \,\Ftot}$ is the reduced DF in $(\Jr,L)$ space\footnote{At $t=0$, since $\Ftot=\Ftot(\Jr,L)$, we have that $F=2L \,\Ftot$. In this paper, at $t=0$, $\Ftot$ is a anisotropic Plummer DF, for some parameter $q$.}. The diffusion coefficients can be explicitly computed from the local velocity deflection coefficients \citep[see equation~3 of][]{Tep2024} 
\begin{equation}
\label{eq:dv_loc_ref}
\begin{bmatrix}
\langle v_{\parallel} \rangle
\\
\langle (v_{\parallel})^2 \rangle
\\
\langle (v_{\perp})^2 \rangle
 \end{bmatrix}
\!=\! 4 \pi m G^{2} \!\ln \Lambda \!\! \int \!\! \rd w \rd \vartheta \rd \phi \sin \vartheta \!
\begin{bmatrix}
- 2 \cos  \vartheta
\\
w \sin^2 \vartheta  
\\
w \, (1+\cos^2 \vartheta)    
\end{bmatrix} \!\Ftot ,\!
\end{equation}
Here, $ \ln \Lambda$ stands for the Coulomb logarithm, for which
we use ${\Lambda\!=\! 0.15 N}$ in the case of single-mass globular clusters following \cite{Henon1975}. From these velocity coefficients, we can then compute the local diffusion coefficients in $E$ and $L$ \citep{BarOr2016}.
\begin{subequations}
\label{eq:nonrotatingdE}
\begin{align}
\dE &= \half \dvParSq +  \half \dvPerpSq + v    \dvPar, \\
\dESq &= v^2 \dvParSq , \\
\dL &= r \frac{\vt}{v} \dvPar + \frac{r^2}{4L} \dvPerpSq,\\
\dLSq &=r^2 \frac{\vt^2}{v^2} \dvParSq + \frac{r^2}{2}   \frac{\vrr^2}{v^2} \dvPerpSq,\\
\dEL &= L \dvParSq.
\end{align}
\end{subequations}
Their orbit-averaging (see Figure~\ref{fig:orbit_average}) then follows from the operation
\begin{equation}
\label{eq:generic_orbit_average}
D_{X} = \frac{\Omega_r}{\pi} \int_{\rrp}^{\rra} \frac{\rd r}{|\vrr|}    \langle \Delta X \rangle (r),
\end{equation}
where $\Omega_r$ is the frequency of radial motion.

\section{Measurements in $N$-body simulations}
\label{app:Nbody_measures}

We proceed mostly as detailed in appendix~{G} of \citetalias{Tep2022}. We generated the initial conditions  from \texttt{PlummerPlus.py}~\citep{Breen2017}. Then, we used the direct $N$-body code \texttt{NBODY6++GPU}~\citep{Wang2015}, version 4.1, to perform the numerical simulations, using the same input file as in appendix~{H1} of~\cite{Fouvry2021}.
Each $N$-body realization was composed of $N=10^{5}$ stars
and integrated up to $\tmax = 1\,000 \, \rHU$
with a data dump every $\Delta t = 1 \, \rHU$. This required about $22 \, \mathrm{h}$ of computation on a 40-core node with a single V100 GPU.
In practice, we considered the set of anisotropic clusters described in Table~\ref{table:para_NBODY}.
\begin{table}[h!]
\centering
\setlength{\tabcolsep}{.25em}
 \begin{tabular}{|c| c|c|c|c|c|c|c|c|c|}
\hline
$q$ & 1 & 0  & -2  & -6  & -16  & -30  \\
\hline
\hline
$\Nrun$ & 100 & 1000 & 100& 200& 100& 100  \\
$(N_{\Jr} , N_{L}) $  & (20,20)& (20,20) & (20,40)& (40,30) &(70,70)  & (70,70)\\
$ \Jr^{\max} $ &  0.55 & 0.55 & 0.55& 0.55&0.55& 0.55\\
$ L^{\max} $&  1.05  & 1.05 &  1.05 & 1.05 &   1.05& 1.05\\
\hline
\end{tabular}
\caption{Detailed parameters for the $N$-body simulations
and the associated binning of action space.
Following equation~\eqref{eq:estimation_F},
we binned the $\bJ = (\Jr,L)$ action space
in $N_{\Jr} \times N_{L}$ uniform bins
within the domain $0 \leq \Jr \leq \Jr^{\max}$
(similarly for $L$).
All quantities are in physical units $G\!=\!M\!=\!b\!=\!1$ if not stated otherwise.
}
\label{table:para_NBODY}
\end{table}

The continued spherical symmetry of the cluster during relaxation (see, e.g., \citetalias{Tep2022}) allows us to compute the instantaneous mean potential, $\psi(r,t)$, using   a simplified approach   based on spherical shells \citep[see, e.g., \S{3} of][]{Henon1971}. Then, for a star with (centered) position and velocity $(\br,\bbv)$,
we compute its specific energy and angular momentum via
\begin{equation}
  E  = \psi(\br) + \frac{\bbv^2}{2},\quad
  L = |\br \times \bbv| .
  \label{eq:Nbody_E_L}
\end{equation}
An estimation of the DF, $F(\Jr,L)$, in the $N$-body runs is obtained by binning the $(\Jr,L)$ action space uniformly (see Table~\ref{table:para_NBODY}).
In particle, for a given action bin of size $\delta \Jr \!\times\! \delta L$
centered around the action coordinates $\bJ = (\Jr , L)$,
we have
\begin{align}
F(\bJ,t) &= \frac{M\, n(\bJ,t)}{(2\pi)^3 \delta \Jr \delta L},
\\
 n(\bJ,t) &= \frac{{\mathrm{stars \ in}} \ [\Jr\!-\!\tfrac{1}{2}\delta \Jr,\Jr\!+\!\tfrac{1}{2}\delta \Jr]\!\times\![L\!-\!\tfrac{1}{2}\delta L,L\!+\!\tfrac{1}{2}\delta L]}{\mathrm{total \ number \ of \ bound \ stars}}. \notag
\end{align}
These time series can be used  to compute the finite difference
\begin{equation}
G(\bJ,T) = \frac{F(\bJ,T) - F(\bJ,0)}{T}.
\label{eq:estimation_F}
\end{equation}
For a carefully chosen $T$, after ensemble-averaging over all available realizations, this yields an (rough) estimation of the rate of change, $\p F/\p t$. However, such an approach suffers from several caveats, which we shall detail in Appendix~\ref{app:interpolation}.

\section{Estimation of $\p F/\p t$ at initial time}
\label{app:interpolation}

\subsection{Polynomial fit}
\label{app:poly_fit}

\begin{figure} 
    \centering
\includegraphics[width=0.45\textwidth]{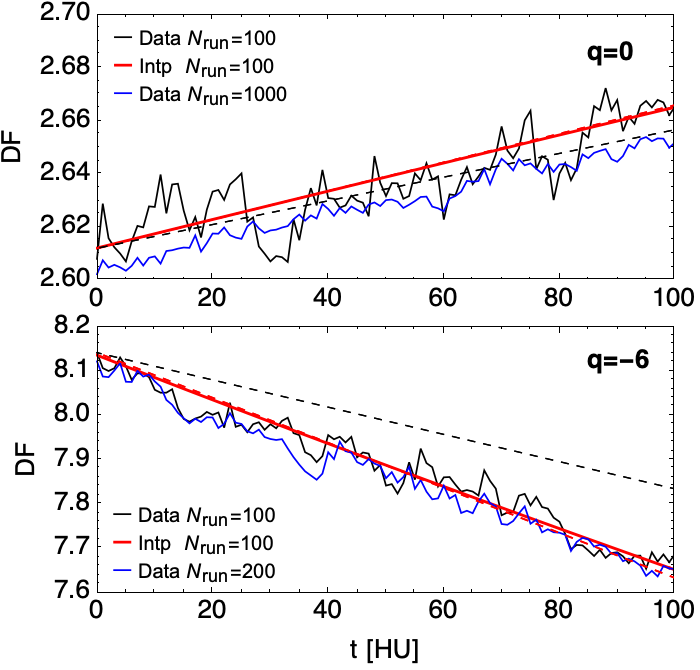}
\caption{Time relaxation of the ensemble-averaged DF (black and blue plots) measured in two sets of $N$-body simulations, at a given action location: for the isotropic case (top panel) $(
   J_{\rm r},L)=(0.041,0.026)$ and for the tangentially anisotropic case $q=-6$ (bottom panel) $(
   J_{\rm r},L)=(0.007,0.718)$. 
   We show, in   solid red, the polynomial time fit of the DF (over 1000 HU). The fit is estimated from the exact DF measurement, in black, averaged over 100 realizations. We show in blue the exact DF, averaged over 1000 runs (resp. 200 runs) for the isotropic (resp. anisotropic) cluster, as a way of   assessing visually the role of fluctuations and the quality of the polynomial fit. 
   Although we observe an overall shift between the blue curve and the red curve   in the upper panel, the slope of the red fit appears to approximately match that of the blue curve. Since we are only interested in that slope to compute $\p F/\p t$ at $t=0$, we conclude that our interpolation scheme is satisfactory. Finally, we compare in dashed lines the finite difference slope (in black) and the fit's slope  at $t=0$ (in red). }
   \label{fig:interpolation_DF}
 \end{figure}

Figure~\ref{fig:interpolation_DF} shows the time evolution of the DF, $F(\Jr,L,t)$,  in a set of $N$-body simulations (in blue),   both for the isotropic cluster and for one anisotropic cluster, for a given set of action variables $(\Jr,L)$.
Its behavior can be decomposed into a smooth, slow evolution, on top of which sharp jumps can be observed. These are the result of the finite $N$-noise of the cluster, and tend to disappear as one ensemble-averages over more realizations. The mean quantity is the ensemble-average of the DF, and is the one whose evolution is described by secular theory. 
Therefore, one needs to have access to this smooth component in order to compute any time derivative.

The most obvious way of doing so is to perform many realizations of the cluster via $N$-body simulations, and to ensemble-average over them. However, this method is very inefficient, as the expected dispersion of the DF, ensemble-averaged over $\Nrun$ realizations, goes as $1/\sqrt{\Nrun}$.
Therefore, we use instead a polynomial interpolation of the DF for each given bin to smooth out the sharpness of the DF's, in the form
\begin{align}
    F(\bJ,t) = \sum_{i=0}^k \alpha_i(\bJ) \,t^i,
    \label{polynomial}
\end{align}
where $k$ is an hyper-parameter one has to set. $k$ should be large enough in order to capture the overall behavior of the DF evolution. However, it should not too high, as this would  lead to  over-fitting and would prevent smoothing out the sharp behavior. We show in red, in Figure~\ref{fig:interpolation_DF}, a polynomial interpolation of the DF for the bins considered. This is fitted to the simulation data in the interval  between $t=0$ and $t=1000 \,\rHU$,  with $k=2$ for $q=0$ and $k=4$ for $q=-6$. 
The initial time derivative therefore reads ${\p F/\p t \!=\! \alpha_1 }$, and is then ensemble-averaged over the cluster's realizations. This yields an estimation of the relaxation rate   which is relatively insensitive to the impact of fluctuations.

\begin{figure*} 
    \centering
\includegraphics[width=0.9\textwidth]{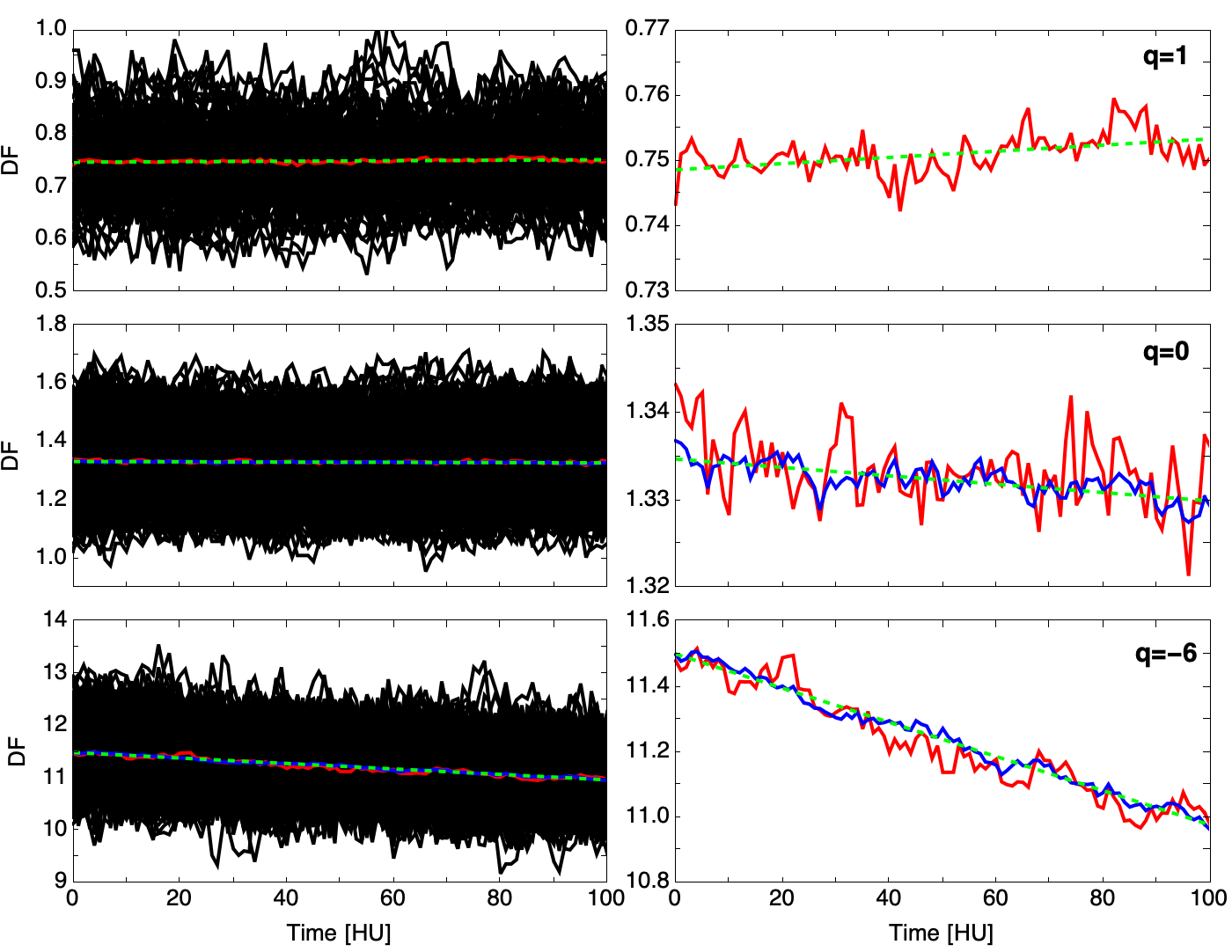}
   \caption{ Illustration of the early time evolution of the DF for three clusters of different anisotropy parameters: a radially anisotropic cluster ($q=1$, top panel $(
   J_{\rm r},L)=(0.041,1.024)$ ), an isotropic cluster ($q=0$, middle panel $(
   J_{\rm r},L)=(0.041,1.024)$ ) and a tangentially anisotropic cluster ($q=-6$, bottom panel $(
   J_{\rm r},L)=(0.021,0.333)$ ). We represent on the left panels how individual realizations can deviate from a mean-field behavior by stacking all our available realizations together in black. We estimate the mean-field DF by taking an ensemble average over 100 realizations (in red), and another one over either 500 ($q=-6$) or 1000 realizations ($q=0$) in blue. Finally, we illustrate the linear regression method,   applied  between $0$ and $t_{\rm end}$ (see Table~\ref{table:ratioNRNBODY_table}), with the dashed green line. The right panels are a zoom-in on the ensemble averages shown in the left panels. Of course, increasing the number of realizations makes the ensemble-average less subject to fluctuations.
}
   \label{fig:interpolation_LR_DF}
 \end{figure*}

Figure~\ref{fig:interpolation_DF} shows that the time evolution of the DF is  nearly linear during early times  -- typically during the first  50 to 100 HUs -- before the non-linearities of its time evolution appear. We illustrate this even further  by showing in Figure~\ref{fig:interpolation_LR_DF} the spreads of the realizations of the DF's time evolution for a few clusters with varying initial anisotropies;   again  we observe   (right panels) that the trend of the early time evolution   of the ensemble average is well captured by a linear fit.  We observed this behaviour at several other locations in phase space in all the models, and for this reason the substantive results reported in Section~\ref{comparison}, including Figure~\ref{fig:dFdt_interp} and Tables \ref{table:ratioNRNBODY_table} and \ref{table:ratioNRNBODY_table_alt}, have been obtained by {\sl linear} regression (i.e. $k = 1$ in Equation~\ref{polynomial}) over the time interval $(0, t_e)$, where $t_e$ is given in Table \ref{table:ratioNRNBODY_table}.  The only locations in phase space where linear interpolation gave unsatisfactory results were those at which $F$ is so small that few particles are present; but the value of $F_t$ at such locations is also small, and they contribute little to the values in Tables \ref{table:ratioNRNBODY_table} and \ref{table:ratioNRNBODY_table_alt}.

\subsection{Systematic calculation of the initial slope and error bars}

Defining $\mathbf{F}=\{F(t_i)\}_i$ the vector of empirical means of the DF for each time, $\boldsymbol{\beta}$ the coefficients of the polynomial fit, $\mathbf{X}=\{t_i^j\}_{ij}$ the matrix of time steps, and $\boldsymbol{\epsilon}$ the noise of the DF at each time steps, we have the linear relation
\begin{align}
    \mathbf{F} = \mathbf{X} \boldsymbol{\beta} + \boldsymbol{\epsilon}.
\end{align}
Following \citet{hansen2002}, the best estimator for the fit parameter is given by the relation
\begin{align}
    \hat{\boldsymbol{\beta} } = ( \mathbf{X}^{\rt}  \mathbf{X})^{-1}  \mathbf{X}^{\rt} \mathbf{F},
\end{align}
where $\mathbf{X}^{\rt}$ is the transpose of $\mathbf{X}$, and its covariance matrix is given by
\begin{align}
   \mathbb{V}(\hat{\boldsymbol{\beta} } ) = ( \mathbf{X}^{\rt}  \mathbf{X})^{-1}  \mathbf{X}^{\rt} \boldsymbol{\Omega}\mathbf{X}( \mathbf{X}^{\rt}  \mathbf{X})^{-1} .
\end{align}
Here, we defined the covariance matrix of the data, $\boldsymbol{\Omega}$, by letting
\begin{align}
    \Omega_{ij}=\mathrm{Cov}[\boldsymbol{\epsilon}]_{ij}=\frac{\sigma^2[F(t_i)] }{N_{\mathrm{run}}}\,\delta_{ij},
\end{align}
where $\sigma^2[F(t_i)]$ are the empirical variances of the DF for each time $t_i$. It follows that the estimation for the time gradient of the DF is given by
\begin{align}
    \frac{\p F}{\p t} = \big[\hat{\boldsymbol{\beta} }\big]_{2} \pm \sqrt{\big[\mathbb{V}(\hat{\boldsymbol{\beta} } )\big]_{22}} .
\end{align}
{}

\section{Chandrasekhar prediction of potential-density evolution}
\label{app:potential_Rc}

We wish to use Chandrasekhar's theory to predict the initial change in potential, $\psi(r)$, and density, $\rho(r)$, in the cluster. To that end, we start from the relation between the density and the DF, $\Ftot(\br,\bbv,t)=\Ftot(\bJ,t)$
\begin{equation}
\rho(\br,t) = \int \rd \bbv \,\Ftot(\br,\bbv,t)= \int \rd \bbv\, \Ftot(\Jr,L,t),
\end{equation}
where $\Jr=\Jr(E[\br,\bbv,t],L[\br,\bbv],t)$ such that
\begin{equation}
\Jr(E,L,t) = \frac{1}{\pi} \int_{\rrp(E,L,t)}^{\rra(E,L,t)} \rd r'\,Q(r',E,L,t),
\end{equation}
with
\begin{equation}
Q(r',E,L,t)^2 = 2(E-\psi[r',t]) - \frac{L^2}{r^{\prime 2}},
\end{equation}
and $Q(\rrp[E,L,t],E,L,t)=Q(\rra[E,L,t],E,L,t)=0$. 
 Furthermore,
\begin{subequations}
\begin{align}
E[\br,\bbv,t] &= \psi(\br,t) + \frac{\bbv^2}{2}= \psi(r,t) + \frac{\vrr^2+\vt^2}{2},\\
 L[\br,\bbv] &= |\br \times \bbv| = r \vt.
\end{align}
\end{subequations}
The time derivative of $\rho(\br,t)$ is therefore given by
\begin{equation}
\dot{\rho}(\br,t) = \int \rd \bbv\, \bigg(\frac{\p \Ftot}{\p t} + \frac{\p}{\p t} [\Jr] \frac{\p \Ftot}{\p \Jr}\bigg),
\end{equation}
where
\begin{equation}
\frac{\p \Ftot(\Jr,L,t)}{\p \Jr} = \frac{\p \Ftot(E,L,t)}{\p E}  \frac{\p E}{\p \Jr}.
\end{equation}
In addition, 
\begin{equation}
\frac{\p}{\p t} [\Jr] =\frac{\p E}{\p t} \frac{\p \Jr}{\p E} + \frac{\p \Jr}{\p t},
\end{equation}
where
\begin{subequations}
\begin{align}
\frac{\p E}{\p t}&=\frac{\p \psi}{\p t}(r,t) ,\\
\frac{\p \Jr}{\p t} &=-  \frac{1}{\pi} \int_{\rrp(E,L,t)}^{\rra(E,L,t)} \frac{\rd r'}{Q(r',E,L,t)}\,\frac{\p \psi}{\p t}(r',t).
\end{align}
\end{subequations}
Therefore
\begin{align}
\label{eq:dot_rho}
\dot{\rho}(\br,t) &\!=\! \int \!\!\rd \bbv  \frac{\p \Ftot}{\p t}\! +\! \frac{\p \psi}{\p t} \! \int \!\!\rd \bbv\,\frac{\p \Ftot}{\p E}  \!+ \!\!\int \!\!\rd \bbv\,  \frac{\p \Ftot}{\p E}  \frac{\p E}{\p \Jr}\frac{\p \Jr}{\p t}   \\
&\!=\! \int \!\!\rd \bbv\,  \frac{\p \Ftot}{\p t} \! +\! \frac{\p \psi}{\p t}   \int\!\! \rd \bbv\,\frac{\p \Ftot}{\p E} \! - \!\int \!\!\rd \bbv\,  \frac{\p \Ftot}{\p E}  \bigg\langle\frac{\p \psi}{\p t}  \bigg \rangle \notag .
\end{align}

\subsection{Matrix method}
Equation~\eqref{eq:dot_rho} mixes the time derivatives of $\psi$ and $\rho$, making it tricky to compute either of them as is. To remedy this difficulty, we introduce a bi-orthogonal basis $\big(\rho^{(p)}[\br],\psi^{(p)}[\br]\big)$ such that
\begin{subequations}
\begin{align}
&\psi^{(p)}(\br)=\int \rd \br'\,U(\br,\br') \rho^{(p)}(\br'),\\
&\int \rd \br\, \psi^{(p)*}(\br) \rho^{(q)}(\br) = -\delta_{pq},
\end{align}
\end{subequations}
with ${U(\br,\br')\!=\!-G/|\br-\br'|}$. We may use for the Plummer cluster the Clutton-Brock basis elements \citep{CB1973}, described in appendix~B1 of \citet{Fouvry2021}. Then, we can decompose the potential and density time  derivatives  by using the expansions \begin{subequations}
\label{eq:expansion_drhodt_dpsidt}
\begin{align}
\dot \rho(\br,t) &= \sum_p \dot{a}_p(t) \rho^{(p)}(\br),\\
\dot \psi(\br,t) &= \sum_p \dot{a}_p(t) \psi^{(p)}(\br).
\end{align}
\end{subequations}
For a spherically symmetric, non-rotating Plummer cluster, we may consider the basis elements of the form $(n,\ell,m)=(n,0,0)$ only, hence 
\begin{subequations}
\begin{align}
\psi^{(p)}(\br) &= \frac{1}{\sqrt{4\pi}} U_{n}^{0}(r),\\
 \rho^{(p)}(\br) &=\frac{1}{\sqrt{4\pi}} D_{n}^{0}(r) .
\end{align}
\end{subequations}
We may then express equation~\eqref{eq:dot_rho} under the form
\begin{align}
\label{eq:dot_ap_sum}
\sum_p \dot{a}_p(t) \rho^{(p)}(\br) 
&= \int \rd \bbv\,  \frac{\p \Ftot}{\p t}  \\
&+ \sum_p \dot{a}_p(t)  \psi^{(p)}(\br)  \int \rd \bbv\,\frac{\p \Ftot}{\p E}  \notag\\
&- \sum_p \dot{a}_p(t) \int \rd \bbv\,  \frac{\p \Ftot}{\p E}  \big\langle \psi^{(p)}  \big \rangle \notag.
\end{align}
We integrate equation~\eqref{eq:dot_ap_sum} against $\psi^{(q)*}(\br)$. This yields
\begin{align}
- \dot{a}_q(t)   
&= \int \rd \br \psi^{(q)*}(\br)\int \rd \bbv\,  \frac{\p \Ftot}{\p t}  \\
&+ \sum_p \dot{a}_p(t) \int \rd \br\, \psi^{(q)*}(\br)\psi^{(p)}(\br)  \int \rd \bbv\,\frac{\p \Ftot}{\p E} \notag \\
&- \sum_p \dot{a}_p(t) \int \rd \br\, \psi^{(q)*}(\br)\int \rd \bbv\,  \frac{\p \Ftot}{\p E}  \big\langle \psi^{(p)}  \big \rangle_{E,L}    ,\notag
\end{align}
where we recall that $E=E(\br,\bbv,t)$ and $L=L(\br,\bbv,t)$. We define the vector/matrix elements
\begin{subequations}
\label{eq:vect_matrix}
\begin{align}
s_q &=-\int \rd \br \psi^{(q)*}(\br)\int \rd \bbv\,  \frac{\p \Ftot}{\p t} ,\\
\mathcal{A}_{q p} &= -\int \rd \br\, \psi^{(q)*}(\br)\psi^{(p)}(\br)  \int \rd \bbv\,\frac{\p \Ftot}{\p E}  ,\\
\mathcal{B}_{q p} &=\int \rd \br\, \psi^{(q)*}(\br)\int \rd \bbv\,  \frac{\p \Ftot}{\p E}  \big\langle \psi^{(p)} \big \rangle_{E,L}  .
\end{align}
\end{subequations}
Therefore, equation~\eqref{eq:dot_ap_sum} takes the matrix form
\begin{equation}
 \dot{\mathbf{a}} = \mathbf{ s} + \mathbf{M} \cdot \dot{\mathbf{a}} ,
\end{equation}
where $ \mathbf{M} = \mathbf{A} + \mathbf{B} $. Upon inversion, this yields
\begin{equation}
 \dot{\mathbf{a}} =\big(\mathbf{I}-\mathbf{M}\big)^{-1} \mathbf{ s}   ,
\end{equation}
where $\mathbf{I}$ is the identity matrix.

\subsection{Action space formulation}

In equations.~\eqref{eq:vect_matrix}, we do not know how to compute \textit{a priori} the 3D relaxation rate, $\p \Ftot/\p t$. We could in theory compute the full 3D FP equation. However, this would introduce unnecessary difficulties. Instead, we wish to use the FP equation involving the reduced DF in the $(\Jr,L)$ space, $\p F/\p t$. We start from the general relation
\begin{equation}
F(\Jr,L) = \int_{-L}^{L} \rd \Lz \,\Ftot \Rightarrow \frac{\p F}{\p t}(\Jr,L) = \int_{-L}^{L} \rd \Lz \,\frac{\p \Ftot}{\p t}.
\end{equation}
To do so, we apply the canonical transformation $(\br,\bbv) \mapsto (\btheta,\bJ)$ in equations~\eqref{eq:vect_matrix}. Therefore, we obtain
\begin{subequations}
\begin{align}
s_q &=-\int \rd \bJ \rd \btheta\, \psi^{(q)*}(r[\theta_1]) \,  \frac{\p \Ftot}{\p t}  ,\\
\mathcal{A}_{q p} &= -\int  \rd \bJ \rd \btheta\, \psi^{(q)*}(r[\theta_1])\,\psi^{(p)}(r[\theta_1])   \,\frac{\p \Ftot}{\p E}  ,\\
\mathcal{B}_{q p} &=\int \rd \bJ \rd \btheta\, \psi^{(q)*}(r[\theta_1]) \,  \frac{\p \Ftot}{\p E}  \big\langle \psi^{(p)}  \big \rangle   .
\end{align}
\end{subequations}
Integrating over $\Lz$ and ordering integration variables appropriately yields
\begin{subequations}
\begin{align}
s_q &=\!-\!\int \rd \Jr \rd L \,  \frac{\p F}{\p t}  \int \rd \btheta\, \psi^{(q)*}(r[\theta_1])  ,\\
\mathcal{A}_{q p} &=\! -\!\int   \rd \Jr \rd L   \, \frac{\p F}{\p E} \int \rd \btheta\, \psi^{(q)*}(r[\theta_1])\,\psi^{(p)}(r[\theta_1])   ,\\
\mathcal{B}_{q p} &=\!\int \rd \Jr \rd L   \,  \frac{\p F}{\p E}  \big\langle \psi^{(p)}  \big \rangle \int\rd \btheta\, \psi^{(q)*}(r[\theta_1])  .
\end{align}
\end{subequations}
The angle integration is the usual orbit-average, which can be done using the radial parameterization or an effective anomaly parameterization
\begin{subequations}
\begin{align}
s_q &=-(2\pi)^3 \int \rd \Jr \rd L \,  \frac{\p F}{\p t}  \big \langle \psi^{(q)*} \big \rangle  ,\\
\mathcal{A}_{q p} &= -(2\pi)^3 \int   \rd \Jr \rd L   \, \frac{\p F}{\p E} \big \langle  \psi^{(q)*} \,\psi^{(p)} \big \rangle  ,\\
\mathcal{B}_{q p} &=(2\pi)^3 \int \rd \Jr \rd L   \,  \frac{\p F}{\p E}  \big \langle  \psi^{(q)*}\big \rangle  \big\langle \psi^{(p)}  \big \rangle .
\end{align}
\end{subequations}
At this point, only three integrations remain, and we made use of the spherical symmetry of the system as well as its non-rotating property.

\subsection{Core radius evolution}
In the continuum limit, the core radius, $\Rc$, is given by
\begin{equation}
\Rc^2 = \frac{\int \rd \br \rho(\br)^3 r^2}{\int \rd \br \rho(\br)^3}.
\end{equation}
Its time derivative is therefore given by
\begin{equation}
 \dot{\Rc} = \frac{3 \int \rd \br \dot{\rho}(\br) \rho(\br)^2 r^2}{2\Rc \int \rd \br \rho(\br)^3} -  \frac{ 3 \Rc \int \rd \br \dot{\rho}(\br) \rho(\br)^2 }{2\int \rd \br \rho(\br)^3}.
\end{equation}
Using equations~\eqref{eq:expansion_drhodt_dpsidt}, we can compute $\dot{\rho}$ from its bi-orthogonal expansion, hence $ \dot{\Rc}$ as well.

\section{Probing the impact of anisotropy on the Coulomb logarithm}
\label{app:Coulomb_parameter_anisotropy}

In this appendix, we study the impact of anisotropy on the Coulomb logarithm $\Lambda$ using first a toy model, and then a Plummer cluster. Following \citet{Henon1975}, the Coulomb parameter takes the form $\Lambda= \gamma N$, where $\gamma$ is a numerical prefactor given by the equation
\begin{align}
\ln \gamma = \ln(0.4) + I_2/I_1,
\end{align}
in the case of equal-mass stars.
Here, we introduced two integrals
\begin{subequations}
\label{eq:coulomb_para_ratio}
\begin{align}
I_1 &= \int \rd \bbv \rd \bbv' \frac{F(\bbv) F(\bbv')}{|\bbv-\bbv'|},\\
I_2 &= \int \rd \bbv \rd \bbv' \frac{F(\bbv) F(\bbv')}{|\bbv-\bbv'|} \ln \bigg(\frac{|\bbv-\bbv'|^2}{2\langle v^2 \rangle}\bigg),
\end{align}
\end{subequations}
where $F(\bbv)$ is the distribution of the background and $\langle v^2 \rangle$ is the mean square velocity.

\subsection{Maxwellian model}

Assuming an isotropic Maxwellian distribution
\begin{align}
F(\bbv) \propto \exp\big[- v^2\big],
\end{align}
\citet{Henon1975} showed that $\gamma=0.1497$ (usually rounded up to $0.15$).

\begin{figure} 
    \centering
   \includegraphics[width=0.45\textwidth]{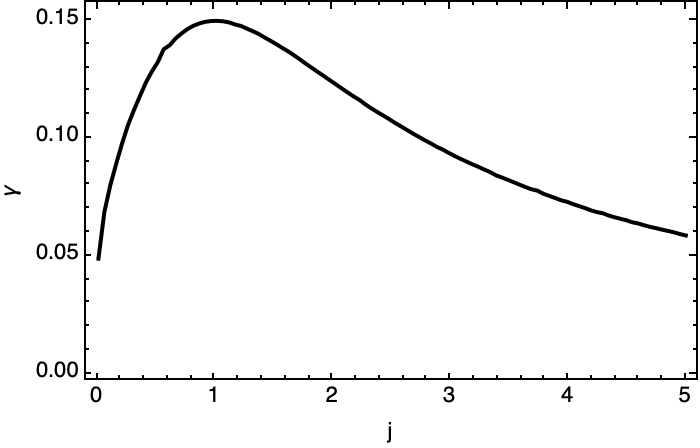}
\caption{ Coulomb logarithm arising from the anisotropic Maxwellian background distribution given by equation~\eqref{eq:aniso_Maxwell}. 
Increasing the anisotropy has the effect of lowering the Coulomb parameter, hence $\ln \Lambda$.
      }
   \label{fig:gamma_j}
 \end{figure}

Let us introduce a degree of anisotropy in the Maxwellian distribution by taking
\begin{align}
\label{eq:aniso_Maxwell}
F(\bbv) \propto \exp\big[- v_1^2 - j^2 \big(v_2^2 + v_3^2\big)\big].
\end{align}
This reduces to the isotropic case when $j=1$. We can numerically evaluate $\gamma$ as a function of $j$, which we show in Figure~\ref{fig:gamma_j}.
Increasing anisotropy decreases the Coulomb parameter, meaning that the value of the anisotropy-dependent Coulomb logarithm decreases as we stray from isotropy.

\subsection{Plummer model}

\begin{figure} 
    \centering
   \includegraphics[width=0.45\textwidth]{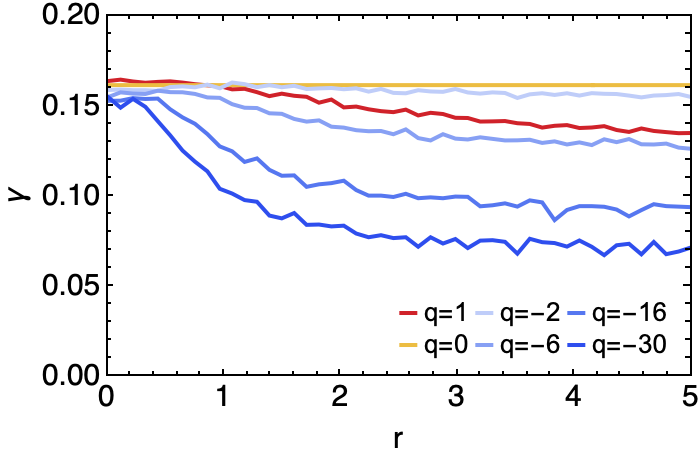}
\caption{ Coulomb logarithm arising from the anisotropic Plummer distribution considered in this paper. 
Equations~\eqref{eq:coulomb_para_ratio} are evaluated using Monte Carlo integration for anisotropic clusters after dealing with the integrable singularities. We observe that $\ln \Lambda$ is mostly maximal for the isotropic cluster.  }
   \label{fig:gammaPlummer_q}
 \end{figure}

We can apply this calculation to the family of anisotropic Plummer clusters studied in this paper. In particular, the mean square velocity is given by \citep{Dejonghe1987}
\begin{equation}
\langle v^2 \rangle(r)=\frac{G M}{6-q} \frac{1}{\sqrt{b^2+r^2}} \bigg(3 -q  \frac{r^2}{b^2+r^2}\bigg).
\end{equation}
This yields a radius-dependent $\gamma$ parameter, which we represent in Figure~\ref{fig:gammaPlummer_q}.
Once again, the $\gamma$ parameter is mostly maximal for the isotropic cluster. As it turns out, it evaluates to 0.162, which is quite close to the Maxwellian value.

As we stray from isotropy, the value of the anisotropy-dependent Coulomb logarithm decreases. If we had used this value instead of the isotropic one, the amplitude mismatch reported in Tables~\ref{table:ratioNRNBODY_table} and \ref{table:ratioNRNBODY_table_alt}  would have been less important.
However, a radial mismatch still remains and cannot be fully explained by the impact of anisotropy alone.

\end{document}